\documentclass[useAMS,usenatbib]{mn2e}
\usepackage{graphicx}
\usepackage{subfigure}

\title[On The Detection of Artefacts in Spectro-Astrometry]{On The Detection of Artefacts in Spectro-Astrometry}
\author[E. Brannigan, M. Takami, A. Chrysostomou and J. Bailey]{E.Brannigan$^{1}$\thanks{E-mail:
e.brannigan@star.herts.ac.uk}, M. Takami$^{1,2}$, A. Chrysostomou$^{1}$, J. Bailey$^{3}$\\
$^{1}$Centre for Astrophysics Research, University of Hertfordshire, Hatfield, HERTS AL10 9AB, UK\\
$^{2}$Subaru Telescope, 650 North A'ohoku Place, Hilo, HI 96720,
USA\\
$^{3}$Anglo-Australian Observatory, PO Box 296, Epping, NSW 1710,
Australia}

\begin{document}

\date{Accepted 2005 ??. Received ????; in original form ????}

\pagerange{\pageref{firstpage}--\pageref{lastpage}} \pubyear{2002}

\maketitle

\label{firstpage}

\begin{abstract}
We demonstrate that artificial bipolar structure can be detected
using spectro-astrometry when the point spread function (PSF) of a
point source suffers distortion in a relatively wide slit.
Spectro-Astrometry is a technique which allows us to probe the
spatial structure of astronomical sources on milliarcseond (mas)
scales making it possible to detect close binaries and to study the
geometry and kinematics of outflowing gas on scales much smaller
than the seeing or the diffraction limit of the telescope. It is
demonstrated that distortion of the PSF, caused by tracking errors
of the telescope or unstable active optics during an exposure can
induce artificial signals which may be misinterpreted as a real
spectro-astrometric signal.  Using simulations we show that these
may be minimised by using a narrow slit relative to the seeing.
Spectra should be obtained at anti-parallel slit position angles
(e.g., $0^{\circ}$ and $180^{\circ}$) for comparison in order to
allow artificial signatures to be identified.
\end{abstract}


\section{Introduction}
Spatial resolutions of optical-IR facilities have dramatically
improved over the last decade.  Indeed, the \emph{Hubble Space
Telescope} and adaptive optics on ground based 8-10m telescopes have
provided spatial resolutions of 0.05 arcseconds, leading to a better
understanding of the nature of a variety of astronomical objects.
Despite this, even higher spatial resolutions are desired to study
the geometry of extrasolar planetary systems, populations and
formation of close binaries, mechanisms of mass ejection and
accretion in young and evolved stars and the nature of active
galactic nuclei. Optical-IR interferometry has begun to provide
resolutions on milliarcsecond (mas) scales, however its
applicability is limited due to the complexity of the technique.

Spectro-astrometry is an alternative approach to study
milliarcsecond structures at optical-IR wavelengths.  The basic
concept is to measure the relative position of the source as a
function of wavelength using either an intensity weighted
centroiding algorithm or profile fitting (see Bailey 1998a for a
more detailed description). The resultant ``position spectrum'' will
show structure on any spectral feature that is displaced from the
centroid of the continuum source: e.g., an emission line arising in
a binary companion, outflowing jets or any other structure that is
not perfectly symmetric about the centroidal source, or whose
displacement from the source varies with wavelength. The smallest
spatial scale observed is limited by the seeing or the diffraction
limit of the telescope, however this technique indeed provides
astronomically useful information at spatial scales much smaller
than these.

The concept was introduced in the 1980's as ``Differential Speckle
Interferometry'' or ``Chromatic Position Difference'' (e.g.,Beckers
1982), and demonstrated to detect close binaries (e.g., Sorokin \&
Tokovinin 1985).  This broad band technique was limited by the
necessity to correct for atmospheric dispersion which can be much
larger than the observed structure.  This difficulty is removed by
measuring the displacement of a spectral line at high resolution.
Aime et al. (1988) were able to measure spectro-astrometric
structure using special instrumentation. See Bailey (1998a) for
review of these works.

More recently Bailey (1998a) adapted this technique using a standard
long-slit CCD spectrograph and coined the term
``spectro-astrometry''.  The method is similar to that used by Solf
\& B\"ohm (1993), who studied a jet from a young stellar object
(YSO) on subarcsecond/arcsecond scales. Bailey (1998a) revised this
method and has shown that with good spatial uniformity of the CCD
coupled with a pixel scale that allows excellent sampling of the
seing profile positional accuracies as small as 1mas are achieved
(see Takami 2003). Bailey has also shown that the technique is a
powerful tool for discovering binary companions toward young and
evolved stars, separating their spectra, and studying the kinematics
of YSO jets/winds and narrow line regions of active galactic nuclei
(AGN). Since this method doesn't require any special
instrumentation, its popularity has steadily increased in recent
years. Studies made for young binaries and YSO jets include
Bailey(1998b), Takami et.al. (2001,2002,2003), Davis et al.
(2001,2003), Whelan, Ray \& Davis(2004) and Baines et al. (2004).
Measurements using an integral field unit (IFU) were first conducted
by Garcia, Thi\'ebaut \& Bacon (1999).  A similar approach using
VLTI-AMBER is investigated by Petrov (2003), Marconi et al. (2003)
and Dougadous et al. (2003). Since the positional accuracy achieved
depends on the PSF size as well as the photon noise, using an
interferometer for this approach has the potential to achieve
information on spatial structures at \emph{micro}arcsecond scales
(Bailey 1998a).

Instrumental effects which may compromise spectro-astrometric
observations have not been fully investigated. These effects may
indeed create false detections of close binaries or bipolar
outflows. Possible instrumental effects include; misalignment of the
spectrum with CCD columns, any departure of the CCD pixels from a
regular grid, imperfect flatfielding or charge transfer deficiencies
in the CCD (Bailey 1998b; Takami et al. 2001). To eliminate
instrumental effects, Bailey (1998a,b) proposed to obtain position
spectra at anti-parallel slit position angles (e.g.,0$^{\circ}$ and
$180^{\circ}$) via rotation of the instrument. On comparison of
these spectra, any real signal from the object will change sign,
while the instrumental effects remain constant.  We can thus
eliminate instrumental effects by subtracting one position spectrum
from the other.  This method is followed by Takami et al.
(2001,2002,2003) and Baines (2004). Garcia et al. (1999), Davis et
al. (2001,2003) and Whelan et al. (2004) however eliminate
instrumental effects by polynomial fitting of the continuum
position.

Bailey (1998a) reports a systematic effect upon observation of sharp
unresolved lines.  This is caused by either telescope tracking
errors or unstable active optics during the exposure, and appears in
the position spectrum as a P Cygni type profile. Bailey warns that
the target spectral lines should be well resolved to observe true
positional displacement.  This instrumental effect could be
eliminated in some cases by Bailey's method, but this cannot be done
by polynomial fitting of the continuum position.  In this paper we
show that this instrumental effect can appear even if the line
profile is fully resolved as was the case in position spectra of RU
Lupi, obtained using the 3.5m New Technology Telescope (NTT). Takami
et al. had previously detected the presence of bipolar outflow in
this object using the 3.9m Anglo-Australian Telescope (AAT). Indeed
the observed positional displacement in the NTT looks similar to
that seen in the AAT data.  We show that the signal detected in the
NTT data is not real, motivating us to investigate the effect of
uneven illumination on spectro-astrometry, how to avoid it and
whether previously detected bipolar outflows are in fact real.

In \S2 we describe the spectro-astrometric observations obtained
using EMMI on the 3.6m NTT and the results in which artificial
signatures were detected. These artefacts are identified by
qualitatively inspecting position spectra obtained using
anti-parallel slit position angles. In \S3 we (1) show that the
observed position spectra are explained by distortion/motion of the
stellar image in a relatively wide slit, (2) show that the extent of
the artefacts, in a given set of observing conditions, can be
greatly reduced by using a slit width narrow in comparison to the
seeing, and (3) show an example of the effect these artefacts have
on real spectro-astrometric data. In \S4 we give some
recommendations on how to monitor the data for the false signatures
of bipolar structure. In Appendix A, we show that the bipolar
outflows reported by Takami et al. (2001,2003) are real and not due
to the instrumental effect described here.

\section[]{Observations and Results}
  \begin{figure*}
   \centering
   \includegraphics[angle=-90,width=1.0\textwidth]{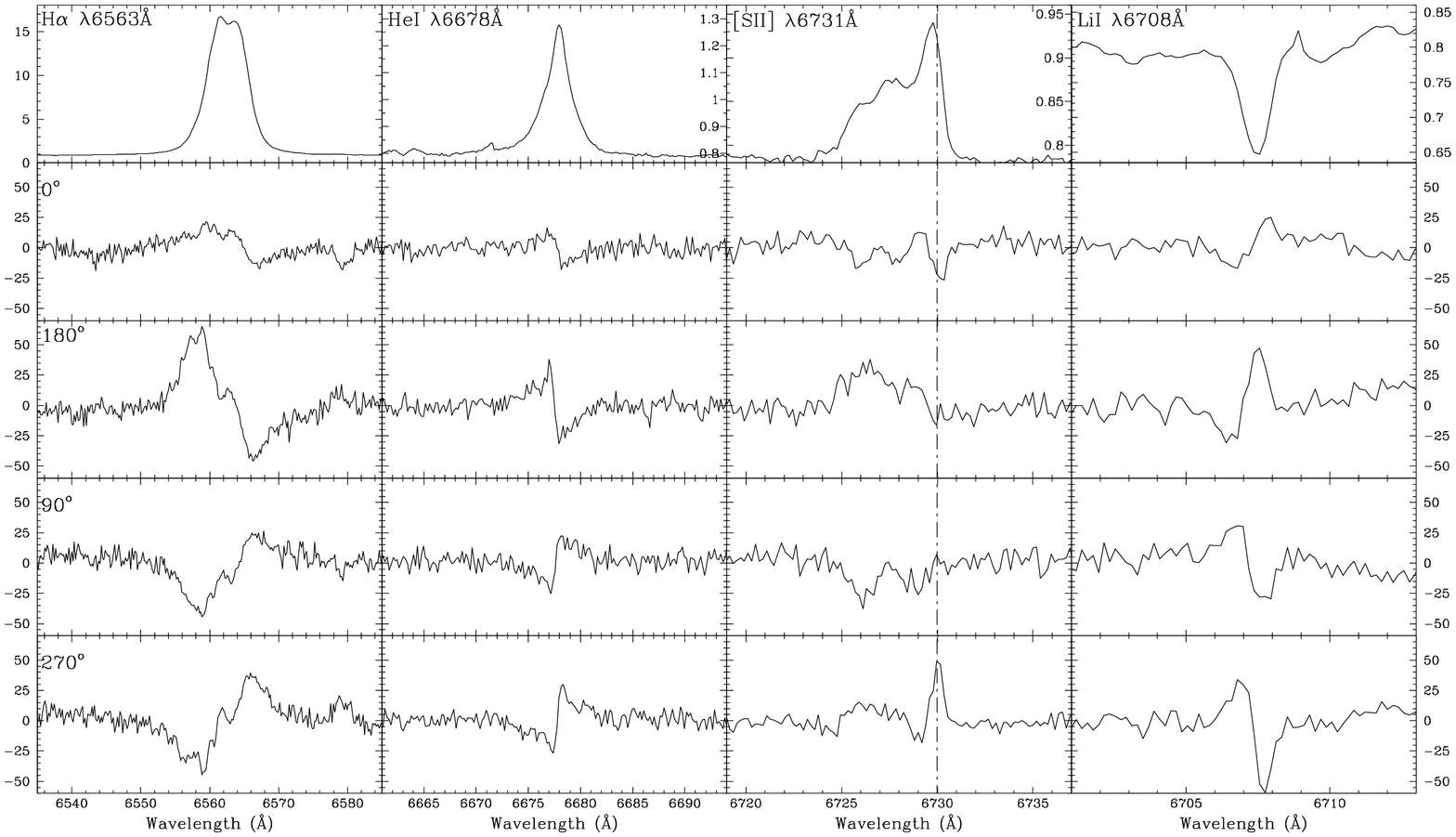}
   \caption{Intensity and position spectra of RU Lup obtained using EMMI-NTT.  From left to
   right;  Top:  H$\alpha$ $\lambda6563$\AA{}, HeI $\lambda 6678$\AA{} \& [SII] $\lambda 6731$\AA{} emission and
   LiI $\lambda 6708$\AA{} absorption lines. The intensities shown are arbitrary;
   Lower four panels:  spectro-astrometric signal (mas) for each corresponding spectral
   feature detected at 4 position angles $0^{\circ},180^{\circ},90^{\circ}\&270^{\circ}$ in the spectrum of RU Lup.
   The bottom two panels show position spectra for anti-parallel slit angles
   $90^{\circ}$ \& $270^{\circ}$ in each of the spectral features,
   similarly in the third and fourth panels from the bottom position spectra for anti-parallel slit angles
   $0^{\circ}$ \& $180^{\circ}$ in each of the spectral features is
   shown. In the blue wing of [SII] $\lambda 6731$\AA{} emission feature
   the angular displacement is seen to change sign in corresponding anti-parallel
   slits, whereas displacement near zero velocity of the [SII] $\lambda 6731$\AA{} line (marked with a dot-dashed line),
   in the permitted emission lines and the LiI $\lambda 6708$\AA{} absorption line does not
   change sign in anti-parallel slits.}
    \label{Fig1}%
    \end{figure*}
Medium resolution spectroscopy of RU Lup was obtained with the 3.5m
NTT in the early morning of April 8 2004, when the seeing was
measured to be 0.8\arcsec. The observations were made using the REad
Medium Dispersion (REMD) mode of the ESO Multi-Mode Instrument
(EMMI) with the \#6 grating and an arcsecond slit. The wavelength
coverage spanned 6100 to 6800\AA{} with resolving \mbox{power
$=5000$}.  The spectra were obtained at 4 position angles:
$0^{\circ}, 90^{\circ}, 180^{\circ}, 270^{\circ}$ by rotating the
instrument. HeAr lamp spectra were used to calibrate the wavelength
scale and a pixel scale of 0.165\arcsec{} with the MIT detector
provided good sampling of the seeing profile.  Flat fields were
obtained by combining many exposures of the spectrograph illuminated
by a halogen lamp.

The data were reduced using the FIGARO package provided by Starlink,
following Bailey (1998a,b) and Takami (2001,2003). After subtracting
the bias level and flat-fielding, ``position spectra'' were
determined by fitting the seeing profile at each wavelength with a
gaussian function.  Global instrumental effects in the position
spectra are removed by polynomial fitting. This allows us to see the
positional displacement of any emission/absorption features at each
slit angle relative to the continuum.  In addition to the position
spectra, the intensity spectra were obtained by subtracting the
bias, flat-fielding, subtracting the adjacent sky and extracting
bright columns on the CCD.

The spectrum of RU Lup shows a wealth of emission lines as observed
by Lago \& Penston (1982) and Takami (2001).  The intensity and
position spectra of the H$\alpha$ $\lambda 6563$\AA{}, HeI $\lambda
6678$\AA{} \& [SII] $\lambda 6731$\AA{} emission lines and LiI
$\lambda 6708$\AA{} absorption line, obtained using EMMI-NTT, are
shown in Figure 1.  The H$\alpha$ emission line shows a large offset
in the position spectrum, $\sim$25mas from the continuum position
and the blue and red shifted wings are displaced in different
directions. Similar displacement in H$\alpha$ is reported in Takami
(2001). Such structure usually indicates the presence of a bipolar
outflow (see Appendix A). However a qualitative inspection of
position spectra, obtained using anti-parallel slits (i.e.
$0^{\circ} \& 180^{\circ}$) in this data, reveals that the
displacement seen in anti-parallel position spectra have equal sign
indicating that the signature must be systematic. In addition, the
displacement observed with slit position angle $0^{\circ}$ is less
than that observed with position angle $180^{\circ}$. In fact two
position spectra observed at a position angle of $0^{\circ}$ exhibit
dissimilar profiles indicating that the effect responsible for the
spurious displacement signal is time variable and as such cannot be
removed even by obtaining position spectra at anti-parallel slit
angles.
    \begin{figure}
        \includegraphics[width=0.5\textwidth]{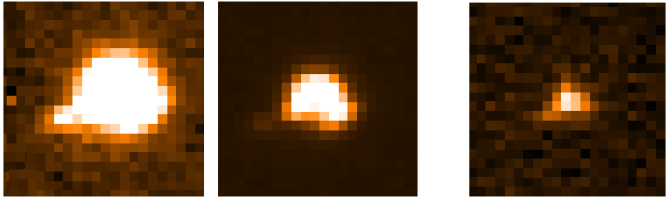}\\
        \caption{Image capture in slit viewer 7/4/2004 using 3.5m NTT.  Pixel scale: 0.4\arcsec.
        Left and Centre: distorted image of the Classical T Tauri Star (CTTS) SZ68. Right: distorted image of the CTTS SZ69}\label{fig2}
    \end{figure}

In addition to H$\alpha$ $\lambda 6563$\AA{} and HeI $\lambda
6678$\AA{}, evidence for bipolar structure is seen in other bright
permitted lines including \mbox{FeII $\lambda
6238/6248/6456/6516$\AA{}}.  The LiI $\lambda 6708$\AA{} absorption
line exhibits displacement similar to that of H$\alpha$ and the
bright permitted emission lines, although in any given spectrum the
direction of displacement in Li I absorption is of opposite sign to
that of the emission lines.  For example, at a position angle of
$180^{\circ}$ the blue and redshifted components of the bright
permitted emission lines show positive and negative displacement
respectively, whereas the Li I absorption line shows negative and
positive displacement in the blue and red wings respectively.

Conversely, the direction of displacement due to blueshifted
forbidden line emission reverses sign at anti-parallel slit
positions ($0^{\circ}$ vs. $180^{\circ}$ ; $90^{\circ}$ vs.
$270^{\circ}$). This is best seen in the [SII] $\lambda 6731$\AA{}
emission line in Figure 1. However a large, sharp displacement is
observed close to zero velocity in the position spectrum at both
[SII] $\lambda 6731$\AA{} and [OI]\protect\footnote{not shown as the
effect is in this line is the same as that in the [SII] line, but
the data at the wavelength of [OI] is of lower quality} $\lambda
6300$\AA{} emission lines. This displacement is observed in all the
position spectra for [SII] $\lambda 6731$\AA{} (marked in Figure 1
with a dot-dashed line) and [OI] $\lambda 6300$\AA{}. This sharp
displacement does not reverse sign in anti-parallel slit position
angles and therefore this is likely to be due to instrumental
effects and not to actual signal from the object.

\section{Induced Signal due to a Distorted Stellar Image in a Wide Slit}
\begin{figure*}
  \includegraphics[angle=-90,width=\textwidth]{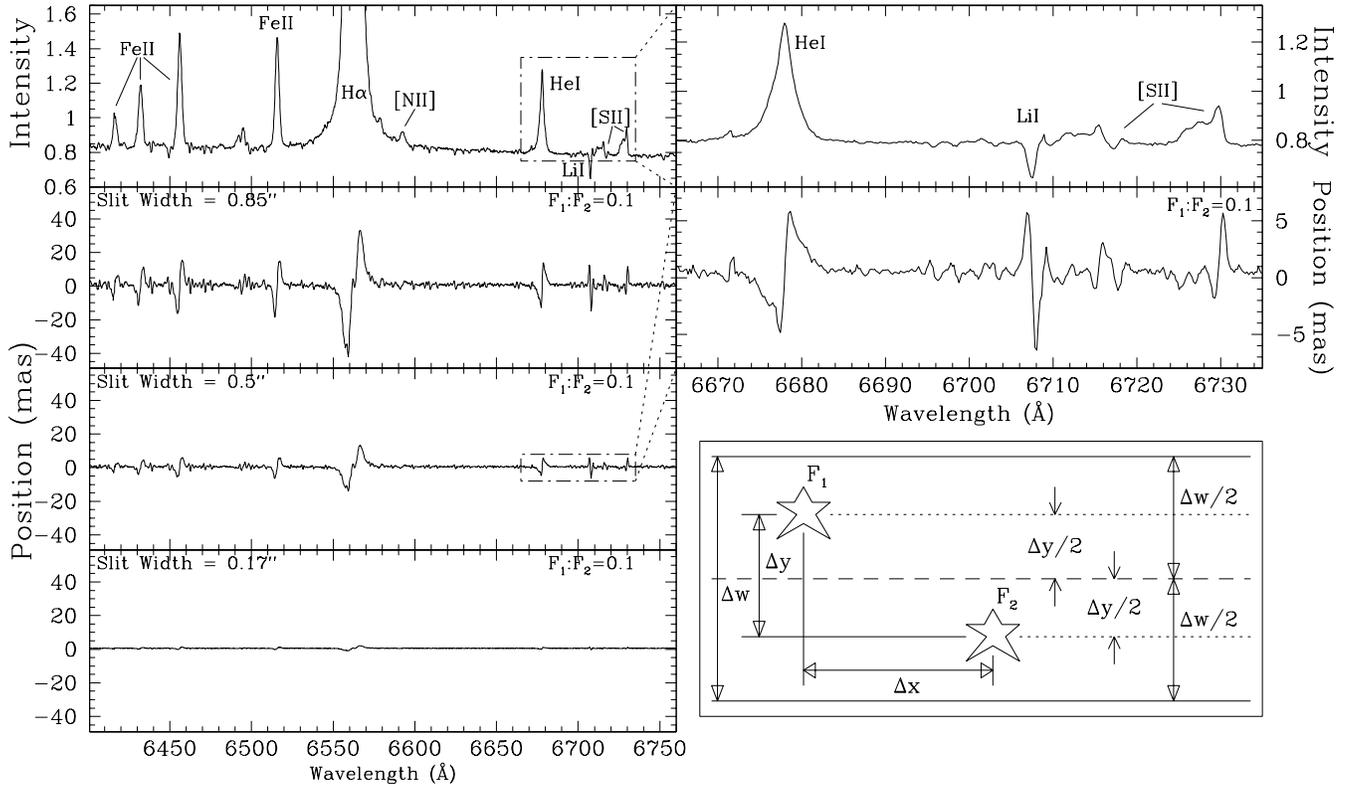}\\
  \caption{Top left: Intensity spectrum of RU Lup obtained on 8/4/2004 using EMMI on the
  3.6m NTT. Lower left: simulated position spectra for a distorted
  PSF using 3 slit widths, 0.85\arcsec{}, 0.5\arcsec{} \&
  0.17\arcsec{}, flux ratio $\textrm{F}_{1}:\textrm{F}_{1}=0.1$
  and two point sources located off centre of the slit with identical spectra.
  Lower right: geometry used in the simulations to show that a distorted PSF can
  reproduce well the artefacts seen in the data. Top right: Close up
  view of the HeI, [SII] and LiI emission and absorption lines.Lower top right:
  positional displacement due to HeI, [SII] and LiI obtained from the
  simulations, where $\textrm{F}_{1}:\textrm{F}_{1}=0.1$ and slit
  width $=0.5$\arcsec{}. The simulated displacement is similar to that seen
  in the data.}
\end{figure*}

The instrumental effects described in \S2 were not observed in
previous observing runs at the AAT described in Takami et al.
(2001,2003). There are a number of key differences between these
earlier observations and those reported in \S2. Primarily, the
typical seeing conditions differ dramatically (0.8\arcsec{} and
1.5\arcsec/2.0\arcsec) while a wide slit of 1.0\arcsec{} was used in
all of the observations.  These artefacts are seen in observations
when the seeing is smaller than the slit width used. Secondly, the
stellar image suffered intermittent distortion during the later
observations, as shown in Figure 2, due to an unstable active optics
system. In \S3.1 it is explained how these combined effects can
induce an artificial signal in the position spectrum, and the extent
of the artefacts that can be expected to occur for a given set of
observing conditions is investigated. In \S3.2 an example of the
effect these artefacts can have on real spectro-astrometric signal
is simulated.

\subsection{Concepts and Simulations}
\begin{table*}
\centering
\begin{tabular*}{\textwidth}{@{\extracolsep{\fill}}r l c c c}\hline
  Parameter & Description & Distortion & Binary & Binary \& Distortion\\\hline
  PS & Pixel Scale & 0.17 & 0.17 & 0.17\\
  $\lambda_{min}$ & start wavelength & 6100 & 6100 & 6100\\
  $\lambda_{max}$ & end wavelength & 6850 & 6850 & 6850\\
  PA$_{bin}$ & Binary position angle & $-$ & 45$^{\circ}$ & 45$^{\circ}$\\
  $\Delta S_{bin}$ & Binary separation & $-$ & 0.1\arcsec{} & 0.1\arcsec\\
  F$_{1}$ & Flux of primary binary component & 1.0 & 3.0 & 3.0\\
  F$_{2}$ & Flux of secondary binary component & $-$ & 1.0 & 1.0\\
  F$_{distort}$ & Percentage of flux in distortion & 0.1 - 0.9 & $-$ & 0.1 - 0.9\\
  PA$_{distort}$ & Distortion position angle & 135$^{\circ}$ & $-$ & 135$^{\circ}$\\
  $\Delta S_{distort}$ & Distance between distortion and ``real
  image'' & 0.1\arcsec{} - 1.0\arcsec{} & $-$ & 0.1\arcsec{} - 1.0\arcsec\\
  FWHM & Gaussian simulated seeing & 0.2\arcsec{} - 1.7\arcsec{} & 0.2\arcsec{} - 1.7\arcsec{}  & 0.2\arcsec{} - 1.7\arcsec\\
  $\Delta w$ & Slit width & 0.1\arcsec{} - 1.4\arcsec{} & 0.1\arcsec{} - 1.4\arcsec{} & 0.1\arcsec{} - 1.4\arcsec\\\hline
\end{tabular*}
\caption{Parameters used in the simulations.}
\end{table*}

When the intensity distribution from a given source is non-uniform
across the slit the resultant spectrum can be slightly blue or
redshifted. This blue/redshifted emission due to ``uneven
illumination'' or ``inaccurate centering'', (e.g., Bacciotti et al.
2002; Ardila et al. 2002) arises from an offset of the light source
from the center of the slit, resulting in a deviation of the angle
of incidence according to d$\alpha$ = d$y$/$f_{col}$, where
$f_{col}$ is the focal length of the collimator. According to the
grating equation:
    \begin{equation}\label{e1}
        m\lambda=(\sin\alpha \pm \sin\beta)
    \end{equation}
where m is the diffracted order, $\lambda$ the diffracted
wavelength, $\alpha$ is the angle of incidence wrt. the normal and
$\beta$ the angle of diffraction wrt. the normal, this leads to a
subsequent deviation in the angle of diffraction which in turn is
translated to a shift in the spectrum. Therefore, the shift in the
spectrum, d$\lambda$, is proportional to the position of the light
source from the center of the slit axis d$y$; d$\lambda = const.
\times$ d$y$. This constant value can be readily determined from the
spectral resolution of the instrument. For the REMD mode of EMMI-NTT
a 1.0\arcsec{} slit provides R$= 5000$, corresponding to $\Delta
\lambda = 1.2$\AA{} giving us the relation:
    \begin{equation}\label{e2}
        \textrm{d} \lambda\; [\AA] = 1.2 \times \textrm{d} y\; [arcsec]
    \end{equation}
As such, if a stellar image is distorted or elongated in the slit,
perhaps due to tracking errors of the telescope or unstable active
optics, the spectrum close to the edges of the slit is slightly blue
or redshifted. This geometry would be replicated by a bipolar
outflow located within the slit.  In order to determine the affect
of these artefacts on spectro-astrometric observations simplified
simulations of two point sources were carried out with their
positions as depicted in Figure 3.  Given that the point sources
have spectra F$_{1}f(\lambda)$ and F$_{2}f(\lambda)$, where F$_{1}$
and F$_{2}$ are constant and $f(\lambda)$ is a normalised spectrum,
the spectra are recorded at the detector as F$_{1}f(\lambda + \Delta
\lambda)$ and F$_{2}f(\lambda - \Delta \lambda)$. Thus the
centroidal position at each wavelength is given by:
    \begin{equation}\label{e4}
    X_{cent}(\lambda) = \frac{\Delta x}{2} \times \frac{\textrm{F}_{1}f(\lambda  +
        1.2 \Delta y) - \textrm{F}_{2}f(\lambda -
        1.2 \Delta y) }{\textrm{F}_{1}f(\lambda + 1.2 \Delta y) +
        \textrm{F}_{2}f(\lambda - 1.2 \Delta y)}
    \end{equation}

To confirm that the artefacts seen in this data are a result of
uneven illumination of the slit simulations of a distorted PSF were
carried out and position spectra obtained using equation (\ref{e4}),
and the geometry described in Figure 3. The simulations reproduce a
number of the artefacts observed in the data (Figure 1),
specifically (see Figure 3): (1) evidence for bipolar structure
exists in the bright permitted emission lines; (2) displacement seen
at the Li I $\lambda 6708$\AA{} absorption line has opposite sign to
that of the emission lines in a given spectrum; and (3) sharp
displacement close to zero velocity is exhibited in the forbidden
emission lines [OI] $\lambda 6300$\AA{} and [SII] $\lambda
6731$\AA{}. The extent of the displacement varies with the ratio
F$_{1}$:F$_{2}$, and in fact the relative displacement in the red
and blue wings in the emission lines in a given position spectrum
also depends upon this ratio. The extent of the displacement
dramatically decreases as the slit width becomes smaller (see Figure
3).

The point source geometry depicted in Figure 3 does not provide a
good representation of observing conditions, as such the seeing
conditions are simulated by convolving the two point sources by use
of a gaussian function. The simulated intensity distribution is then
described as:
  \begin{equation}
    \textrm{F}(x,y) = \textrm{G}_{1}(x,y) + \textrm{G}_{2}(x,y)\label{e5}\\
  \end{equation}
\noindent where:
  \begin{eqnarray}
      \displaystyle \textrm{G}_{1}(x,y) = \frac{2.773}{(FWHM)^{2} \pi} \; \textrm{F}_{1}
      \; e^{-2.773 \frac{(x - \Delta x_{1})^{2} + (y - \Delta
      y_{1})^{2}}{(FWHM)^{2}}}\nonumber\\
      \displaystyle \textrm{G}_{2}(x,y) = \frac{2.773}{(FWHM)^{2}\pi} \; \textrm{F}_{2}
      \; e^{-2.773 \frac{(x - \Delta x_{2})^{2} + (y - \Delta
      y_{2})^{2}}{(FWHM)^{2}}}\nonumber
  \end{eqnarray}
where FWHM is the full width half maximum of the seeing profile,
($\Delta$ x$_{1}$, $\Delta$ y$_{1}$) is the position of the primary
point source, ($\Delta$ x$_{2}$, $\Delta$ y$_{2}$) the position of
the secondary point source, F$_{1}$ and F$_{2}$ are the flux of
primary and secondary point sources respectively. Combining
equations (\ref{e4}) \& (\ref{e5}) results in:
  \begin{equation}\label{e6}
    X_{cent}(\lambda) = \frac{\int\int x \: \textrm{F}(x,y) \: f(\lambda +
    \Delta \lambda(y)) \: \textrm{d}x \: \textrm{d}y}{\int\int \: \textrm{F}(x,y)
     \: f(\lambda + \Delta \lambda(y)) \: \textrm{d}x \: \textrm{d}y}
  \end{equation}

The two point sources, having being allocated the relevant relative
flux, are convolved by our simplified model of the seeing and
subsequently shifted according to equation \ref{e2}. A position
spectrum is then obtained as described by equation \ref{e6}

\begin{figure}
  \includegraphics[height=0.5\textwidth,angle=-90]{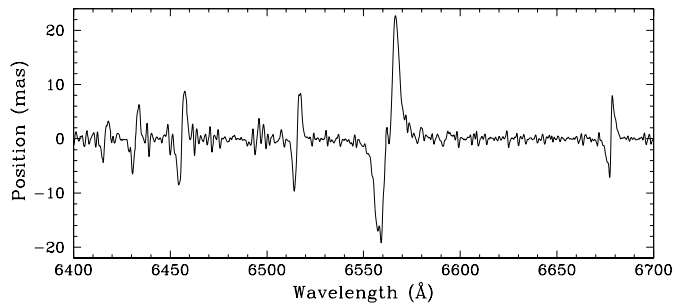}\\
  \caption{Position spectrum obtained as described in \S3.1, where
  the parameters slit width, seeing
  $\Delta$S$_{distort}$ and F$_{distort}$ are 1.0\arcsec{}, 0.8\arcsec{},
  0.7\arcsec{} and 0.5 respectively.}\label{fig5}
\end{figure}

\begin{figure*}
\begin{minipage}{\textwidth}
  \subfigure{
  \label{fig6a}
  \includegraphics[height=0.5\textwidth,angle=-90]{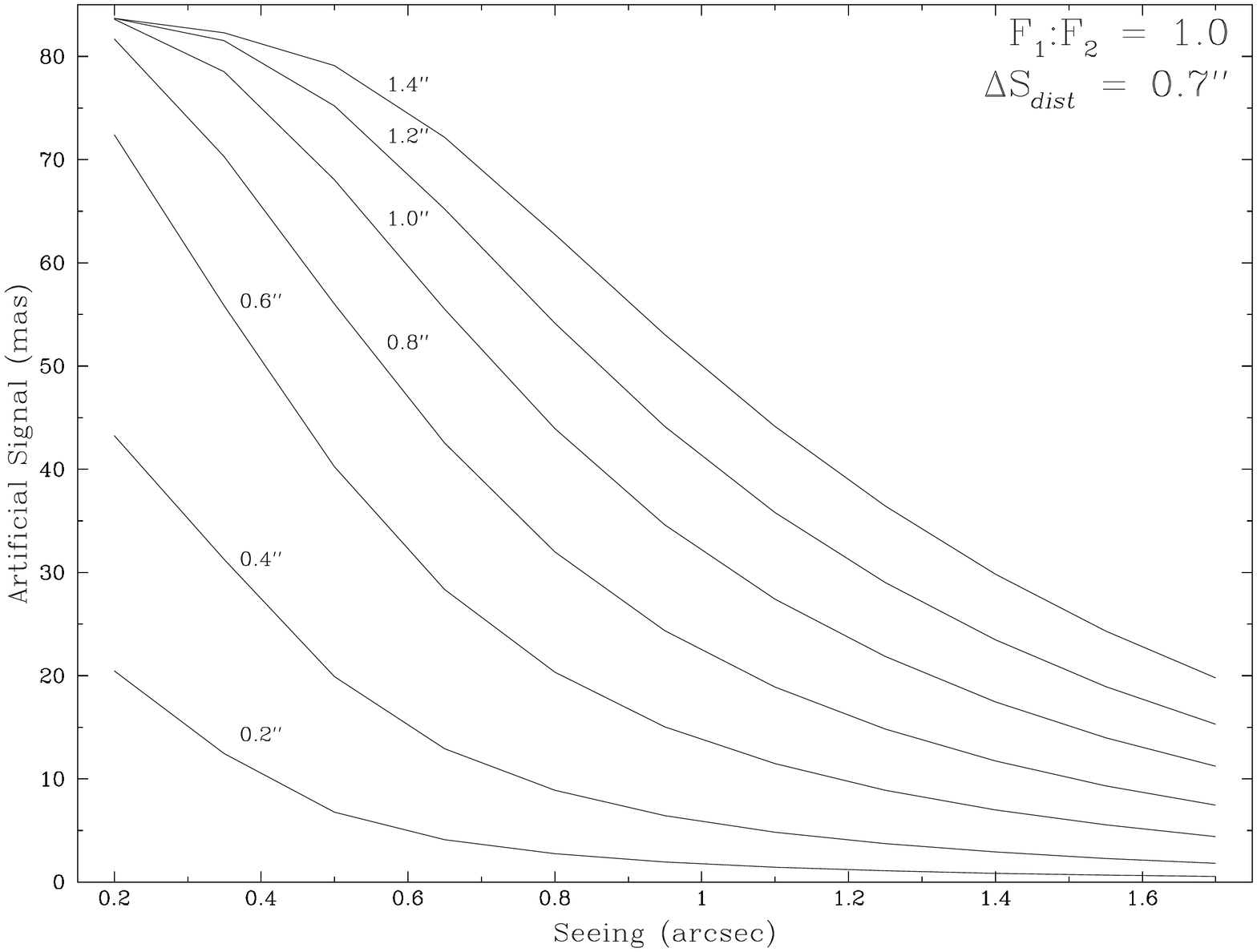}
  }
  \subfigure{
  \label{fig6b}
  \includegraphics[height=0.5\textwidth,angle=-90]{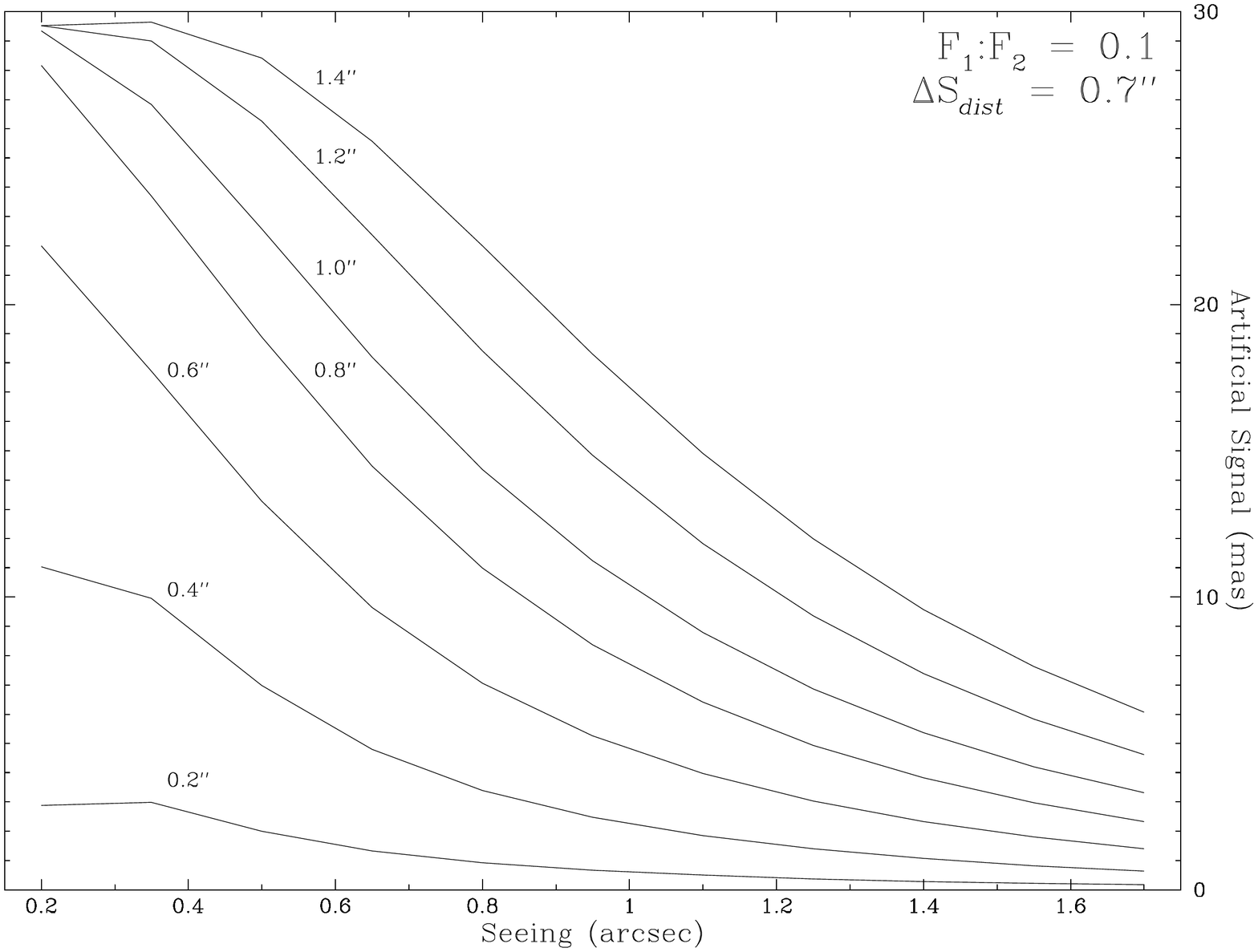}
  }
  \subfigure{
  \label{fig6c}
  \includegraphics[height=0.5\textwidth,angle=-90]{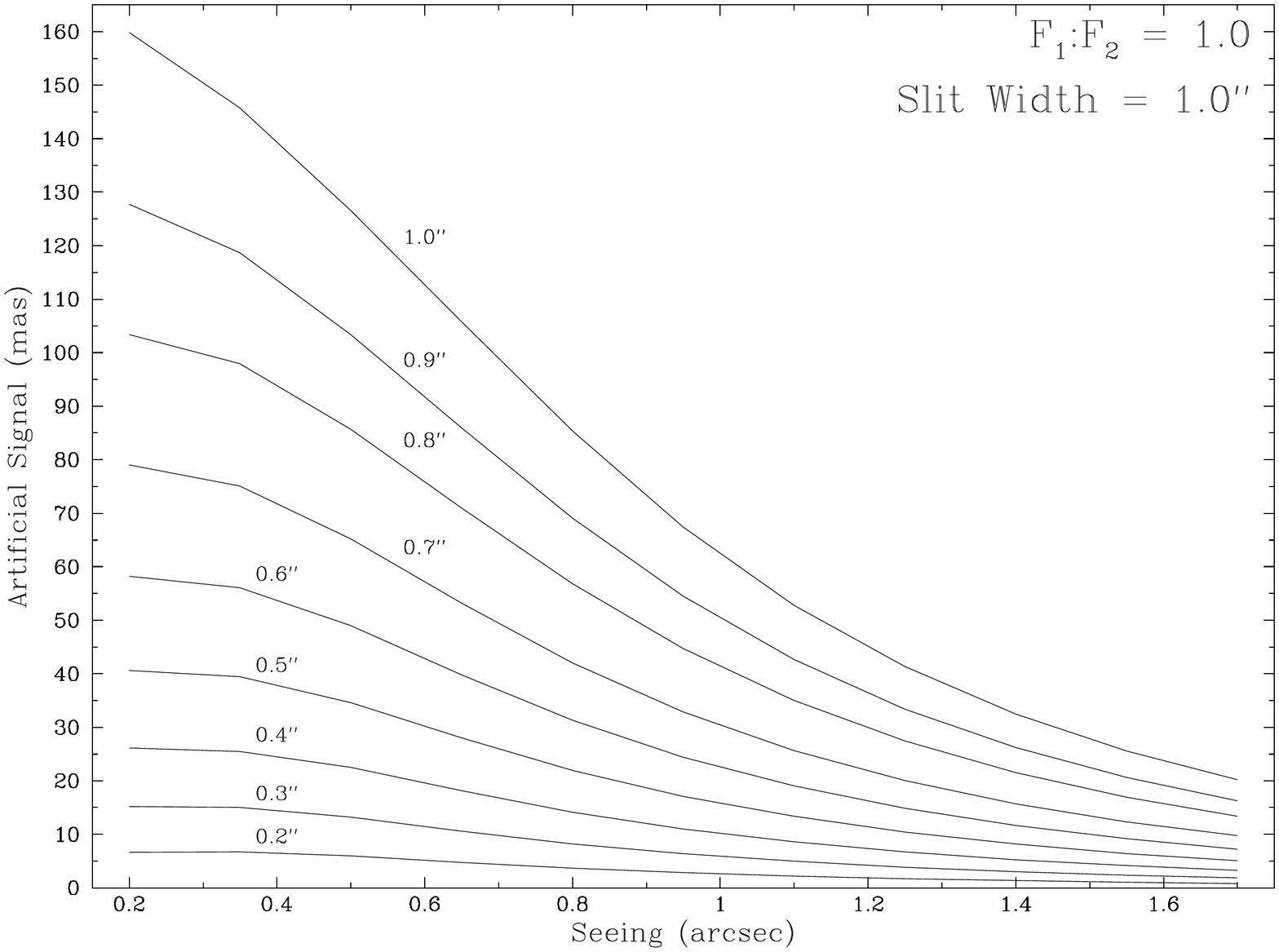}
  }
  \subfigure{
  \label{fig6c}
  \includegraphics[height=0.5\textwidth,angle=-90]{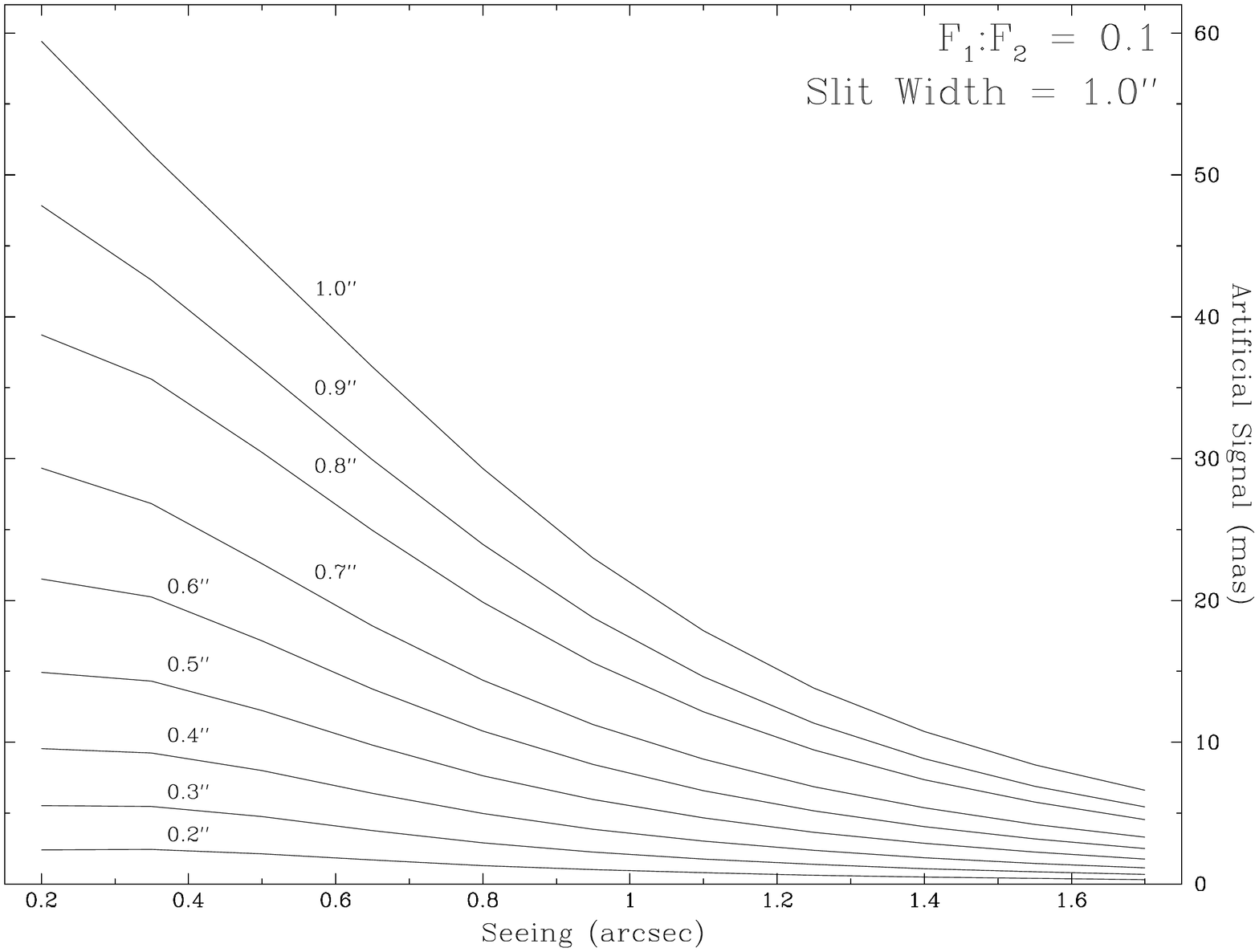}
  }
\caption{Simulated angular displacement in H$\alpha$, measured from
peak to trough, for a range of seeing, slit width, distance between
the ``real image'' and distortion ($\Delta$S$_{distort}$) and the
amount of flux attributed to the distortion (F$_{distort}$, which
gives the ratio F$_{1}$:F$_{2}$), the values of which are given in
Table 1. Top:  Artificial signal generated for a given slit width
(solid lines) for a range of seeing values, with F$_{1}$:F$_{2}$=1.0
(left) and F$_{1}$:F$_{2}$=0.1 (right). Bottom:  Artificial signal
generated for a given separation between ``real image'' and
distortion, $\Delta$S$_{distort}$ (solid lines) for a range of
seeing values, with F$_{1}$:F$_{2}$=1.0 (left) and
F$_{1}$:F$_{2}$=0.1 (right).}
\end{minipage}
\end{figure*}

A simulated position spectrum (Figure 4), obtained as described
above, shows displacement on a scale smaller than those artefacts
obtained by the unconvolved point source. This is because much of
the light falls close to and on the center of the slit and as such
suffers less of a ``shift''. In this instance a 1.0\arcsec{} slit is
simulated with FWHM of 0.8\arcsec{}, as was the case for those
observations shown in Figure 1.  The two point sources were defined
0.7\arcsec{} apart, with equal flux (i.e. F$_{1}$:F$_{2} = 1$) and
both off center of the slit, as described in Figure 3. The angular
scale of the displacement simulated for H$\alpha$ emission
($\sim25$mas) compares well with the artefacts observed in the
position spectra obtained at the NTT.

In order to investigate the limitations imposed on
spectro-astrometry due to systematic effects, the extent of the
artefacts that can be expected when uneven illumination of the slit
occurs (perhaps due to tracking errors or instrumental concerns) for
a given set of conditions is simulated. A single point source is
allocated at the centre of the slit with flux F$_{1}$ and
subsequently convolved by use of a gaussian function. A distortion
in the resulting PSF is simulated by ``relocating'' a percentage of
the flux attributed to the point source (F$_{distort}$) to a new
position off centre of the slit. The results of these simulations
are shown in Figure 6, in which the full extent of the artificial
signal, measured from peak to trough, is shown for a range of
seeing, slit width, distance between the ``real image'' and
distortion ($\Delta$S$_{distort}$) and the amount of flux attributed
to the distortion (F$_{distort}$), the values of which are given in
Table 1. The results are shown for an F$_{distort}$ of 0.1 and 1.0
for given slit widths and $\Delta$S$_{distort}$.  From this we can
clearly see that for small $\Delta$S$_{distort}$ where the
distortion in the PSF cannot be detected by eye artificial
spectro-astrometric signal is still generated on a scale
\mbox{$\sim5$mas} in H$\alpha$, above the typical detection limit of
1mas for previous spectro-astrometric observations (Takami 2001,
2003).  In fact when the input parameters are similar to the
conditions during earlier observations described in Takami 2001,
2003 (i.e. seeing ~1.5\arcsec coupled with a slit width of
1.0\arcsec) the displacement in H$\alpha$ shows an angular scale of
\mbox{$\sim$7 - 8mas}, similarly above the typical detection limit.
Therefore it is imperative to obtain data at anti-parallel slit
angles in order to determine, by qualitative inspection, if any
detected signal is in fact from a real structure in the object, even
when the PSF appears symmetric.

\begin{figure*}
\begin{minipage}{\textwidth}
  \subfigure{
  \label{fig7a}
  \includegraphics[angle=-90,width=0.5\textwidth]{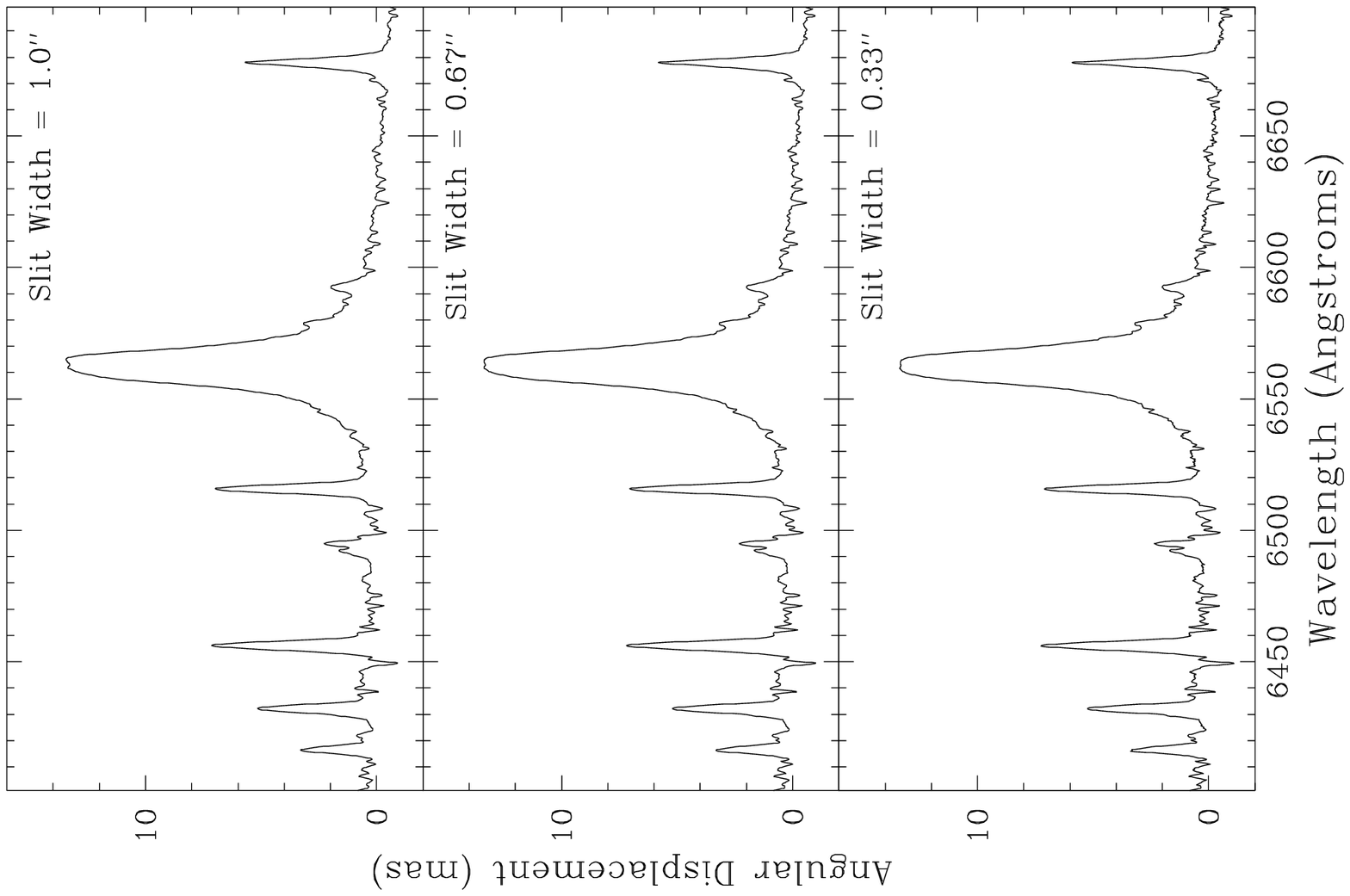}
  }\hspace{0.3cm}
  \subfigure{
  \label{fig7b}
  \includegraphics[angle=-90,width=0.5\textwidth]{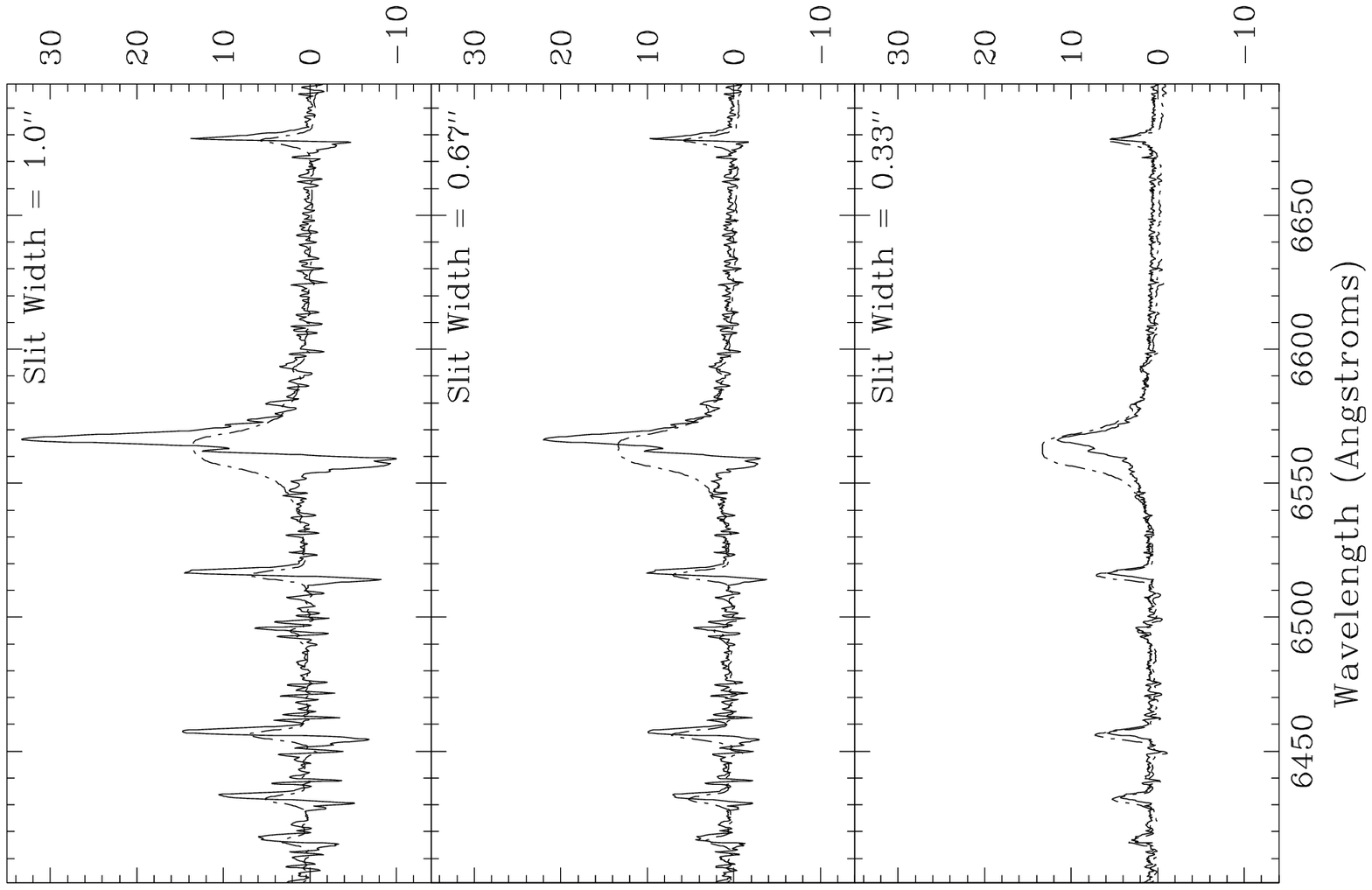}
  }
\caption{Binary simulations with (right) and without (left) a
distorted PSF. The true displacement is represented by a dot-dashed
line in the left panel. A range of slit widths are displayed in each
case: 1.0\arcsec{},
  0.67\arcsec{} \& 0.33\arcsec{} (top to bottom respectively).}
\end{minipage}
\end{figure*}

The model suggests that the effect can be minimised by defocusing
the telescope. For example, Figure 5 shows that given a slit width
of 1.0\arcsec{} an increase of 50\% in the seeing parameter (from
0.8\arcsec{} to 1.2\arcsec) results in a reduction of the angular
scale of the simulated artefacts by a factor of 2.  However, the use
of a slit width narrow in comparison to the seeing reduces the
extent of the artefacts to a greater extent. From Figure 5 it can be
seen that given a seeing of 0.8\arcsec{} a decrease in the slit
width by 50\% (from 1.0\arcsec{} to 0.5\arcsec) results in a
reduction of the angular scale of the simulated artefacts by a
factor of 3.  As such it is clear that a reduction in the slit width
is the best option in order to minimise the magnitude of these
artefacts.

\subsection{Binary Simulations}

In order to determine (1) what, if any, effect the use of a
relatively narrow slit might have on the detection of any real
spectro-astrometric signal from the object, and (2) how this
systematic effect will affect real signals from the object, the
convolved model was adapted with the introduction of an artificial
binary companion. The intensity spectrum allocated to the secondary
binary component is the same as that in the primary with line to
continuum ratio increased by a factor of 2 in order to allow a
``real'' spectro-astrometric signal to be detected in the simulated
binary object, the parameters used in these simulations are listed
in Table 1. A percentage of the binary flux is allocated to a
distortion in the PSF to simulate the effect of ``uneven
illumination''.

Figure 6 shows an example result of the simulations when no
distortion is present, the parameters for the position spectra shown
are those given in Table 1, the specific values of F$_{distort}$,
$\Delta$S$_{distort}$ and seeing (FWHM) are 0.5, 0.7\arcsec{} and
0.8\arcsec{} respectively. From this figure we can see that the
``real'' spectro-astrometric signal is unaffected by a reduction in
slit width, therefore the use of a narrow slit has a minimal effect
on the extent of the real signal from the object and must therefore
be an efficient method of reducing the artificial displacement in
the observed position spectra.

Also shown in Figure 6 is both the spectro-astrometric signal
detected when the PSF suffered distortion, and the ``real''
spectro-astrometric signal that is detected when the PSF suffers no
distortion (dot-dashed line). The simulations show that the extent
of the detected artefacts is reduced by a factor of $\sim1.5$ when
the slit width is reduced by a third (from 1\arcsec{} to
0.67\arcsec{}), and when the slit width is reduced by two thirds
(from 1\arcsec{} to 0.33\arcsec{}) the artefacts are reduced by a
factor of $\sim3$. Caveat: the use of a slit width smaller than the
seeing will result in the loss of photons. As such, the measured
positional accuracy is affected by the reduction in slit width (see
Takami 2003).  For example, if the slit width is reduced by a factor
of 2 the number of photons detected (N) will reduce the signal to
noise ($\propto N^{2}$) by a factor of 4, resulting in a reduction
in the detection limit ($\propto N^{-1/2}$), and hence the
positional accuracy achieved, by a factor of 2.

\section{Conclusions and Recommendations}
Artificial spectro-astrometric signal can be produced in a position
spectrum as a result of a distorted stellar image in a relatively
wide slit, as was the case in data obtained using the 3.6m NTT. The
observed artificial displacement looks similar to that produced when
observing an object exhibiting bipolar structure, the only
difference being that a real signal from bipolar structure will
change sign when observed using anti-parallel slit position angles
whereas the artificial signal will not.

Time variable distortion of the stellar image is a significant
problem, the time variation seen is presumably due to variation of
the PSF, and as such the resulting magnitude of the effect is
time-averaged. As a result the artefacts cannot be removed by
subtracting position spectra obtained using long slit spectra with
anti-parallel slit position angles, as is the case for other
instrumental effects. Spectro-astrometric data must therefore be
carefully monitored during observations to determine when these
artefacts arise and the observations adjusted to compensate.

Simulations show that the use of a slit width narrow in comparison
to the seeing greatly reduces the extent of this instrumental
effect.  However, an image of high quality and good tracking of the
telescope are mandatory in order to obtain position spectra with
high positional accuracy.

Based on observations and simulations we recommend that in order to
minimise these artefacts and obtain position spectra with high
positional accuracy the following are taken into account:
\begin{enumerate}
    \item When observing use a slit width narrow in comparison to
    the seeing. This systematic effect was not observed in
    observations where the seeing is typically 1.5\arcsec{}     - 2.0\arcsec{}
    and the slit width used was 1.0\arcsec.
    \item The use of an integral field unit (IFU) may be
    advantageous as they do not experience uneven illumination (e.g.
    IFU's with micro lenses or fibres), however although previous
    work by Garcia et al. (1999) appear to have been successful,
    instrumental effects arising in IFU's may well be more complicated
    than those from long slit spectrographs.  These should be investigated.
    \item Obtain spectra with a wide spectral range.  The
    systematic effect will create the same artificial signal in a
    number of emission and absorption features.  Therefore if a wide
    spectral range is observed to include a number of absorption and emission
    lines with a range of excitations it will be possible to deduce
    if the bipolar signal is real.
    \item Spectra should be obtained at anti-parallel slit position
    angles.  Not all instrumental effects are well understood or
    anticipated, as such comparison of position spectra obtained
    using anti-parallel slit position angles is the best method to
    determine whether or not the signal detected is due to real
    structure in the source.
\end{enumerate}

\section*{Acknowledgments}
We acknowledge the data analysis facilities provided by the Starlink
Project which is run by CCLRC on behalf of PPARC.  EB and MT thank
PPARC for support through a post-grad studentship and PDRA
respectively.  This research has made use of the NASA's Astrophysics
Data System Abstract Service.

\appendix

\section[]{Previous Results Reported by Takami et al. (2001, 2003)}

\begin{figure*}
\begin{minipage}{\textwidth}
  \includegraphics[width=\textwidth]{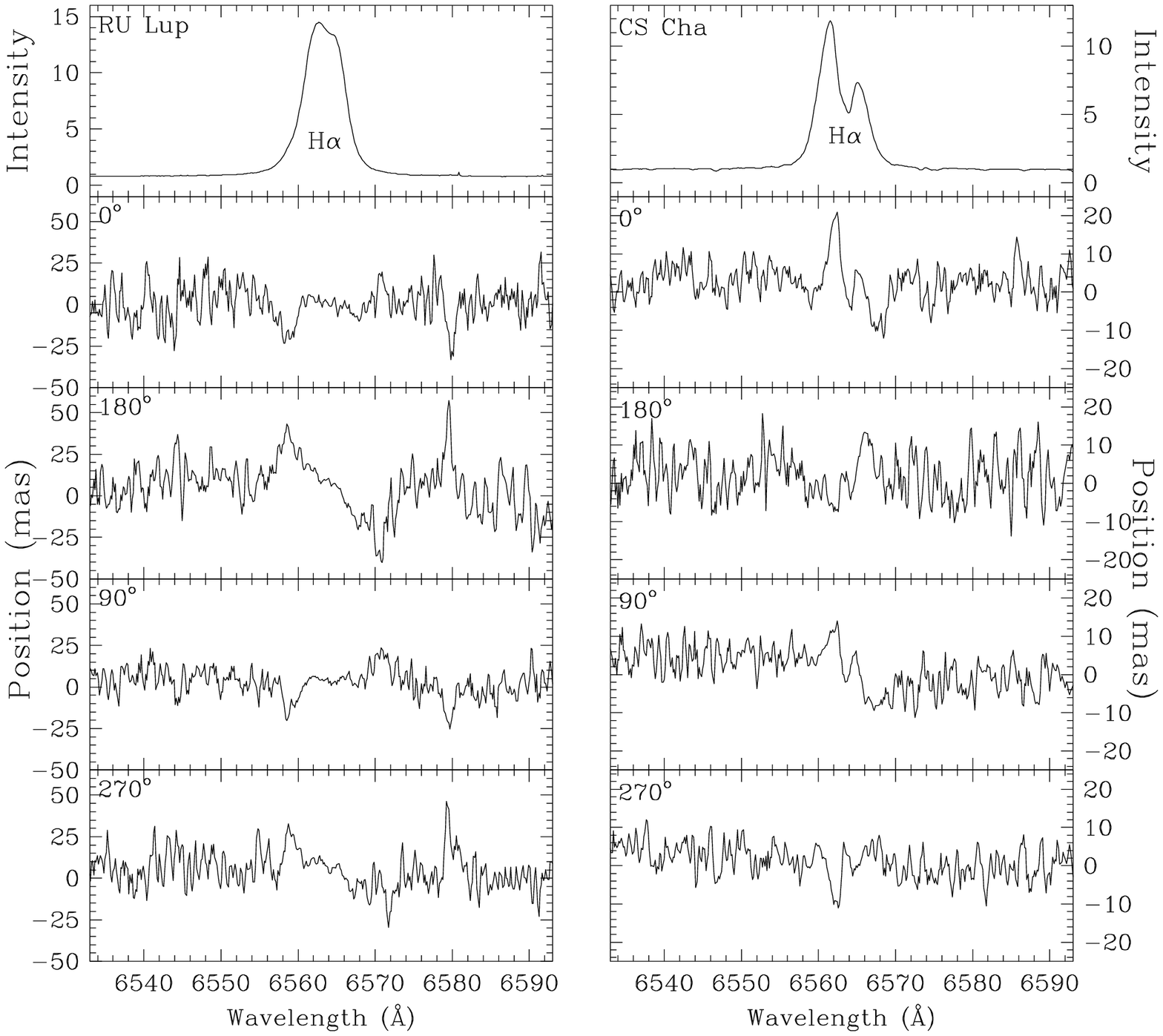}
  \caption{Intensity and position spectra (H$\alpha$) of RU Lup (left) and CS Cha (right)
  reported in Takami (2001) and Takami (2003) respectively.
  Position spectra for each object were obtained at position angles of 0$^{\circ}$,
  90$^{\circ}$, 180$^{\circ}$ \& 270$^{\circ}$.}
\end{minipage}
\end{figure*}

The detection of bipolar outflows in H$\alpha$ has been reported
previously in RU Lup and CS Cha, \cite{b22,b21} respectively, see
Figure A1.  In order to confirm that the results reported are due to
real signals from the objects and not to the artefacts described in
this work the (previously unpublished) anti-parallel slit data,
which do not show the effect, are included in Figure A1.

Displacement from the centroidal continuum position is seen in both
objects at H$\alpha$ $\lambda$6563\AA{}, and additionally at [SII]
$\lambda$6716,6731\AA{} and [NII] $\lambda$6584\AA{} in RU Lup.
Figure A1 clearly shows that spectro-astrometric signal detected in
anti-parallel slits ($0^{\circ}-180^{\circ}$,
$90^{\circ}-270^{\circ}$) are displaced in opposing directions
indicating a genuine signal from the object.  In addition, the
blueshifted H$\alpha$ wing in RU Lup is displaced in the same
direction as the forbidden emission, a well known probe for
outflowing gas. The position angle of the displacement for H$\alpha$
emission for CS Cha is perpendicular to the optical continuum
polarisation, which is very often perpendicular to the jet axis
(\cite{b21}). We thus conclude that the results reported by
\cite{b22,b21} are real and not artefacts.

\bsp \label{lastpage}

\end{document}